\documentclass[amssymb,prb,
twocolumn,
,citeautoscript,floatfix,
footinbib,superscriptaddress]{revtex4}
\usepackage{graphicx}
\usepackage[latin1]{inputenc}
\usepackage{times}
\usepackage{mathptm}
\newcommand\beq{\begin{equation}}
\newcommand\eeq{\end{equation}}
\newcommand\bea{\begin{eqnarray}}
\newcommand\eea{\end{eqnarray}}

\newcommand\ben{\begin{enumerate}}
\newcommand\een{\end{enumerate}}

\newcommand\TTS{Sr$_3$Ru$_2$O$_7$}
\newcommand\TOF{Sr$_2$RuO$_4$}
\newcommand\FTT{Sr$_4$Ru$_3$O$_{10}$}
\newcommand\OOT{SrRuO$_3$}

\begin{document}

\title{\sffamily Quantum oscillations near the metamagnetic transition in {Sr$_3$Ru$_2$O$_7$}}

\author{J.-F. Mercure}
\affiliation{Scottish Universities Physics Alliance (SUPA), School of Physics and Astronomy, University of St Andrews, North Haugh, St Andrews KY16 9SS, United Kingdom}

\author{A. W. Rost}
\affiliation{Scottish Universities Physics Alliance (SUPA), School of Physics and Astronomy, University of St Andrews, North Haugh, St Andrews KY16 9SS, United Kingdom}

\author{E. C. T. O'Farrell}
\affiliation{Cavendish Laboratory, University of Cambridge, J.J. Thomson Avenue, Cambridge CB3 0HE, United Kingdom}

\author{S. K. Goh}
\affiliation{Cavendish Laboratory, University of Cambridge, J.J. Thomson Avenue, Cambridge CB3 0HE, United Kingdom}

\author{R. S. Perry}
\affiliation{SUPA, School of Physics, University of Edinburgh, Mayfield Road, Edinburgh EH9 3JZ, United Kingdom}

\author{M. L. Sutherland}
\affiliation{Cavendish Laboratory, University of Cambridge, J.J. Thomson Avenue, Cambridge CB3 0HE, United Kingdom}

\author{S. A. Grigera}
\affiliation{Scottish Universities Physics Alliance (SUPA), School of Physics and Astronomy, University of St Andrews, North Haugh, St Andrews KY16 9SS, United Kingdom}
\affiliation{Instituto de F\'isica de liquidos y sistemas biologicos, UNLP, 1900 La Plata, Argentina.}

\author{R. A. Borzi}
\affiliation{Instituto de Investigaciones Fisicoqu\'imicas Te\'oricas y Aplicadas (UNLP-CONICET), 
c.c. 16, Suc. 4 and Departamento de F\'isica, IFLP, UNLP, c.c. 
67, 1900 La Plata, Argentina.}
 
\author{P. Gegenwart}
\affiliation{I. Physik. Institut, Georg-August-Universit$\ddot{a}$t G$\ddot{o}$ttingen, D-37077 G$\ddot{o}$ttingen, Germany}

\author{A.S. Gibbs}
\affiliation{Scottish Universities Physics Alliance (SUPA), School of Physics and Astronomy, University of St Andrews, North Haugh, St Andrews KY16 9SS, United Kingdom}
\affiliation{School of Chemistry, University of St Andrews, North Haugh, St Andrews KY16 9ST, United Kingdom}

\author{A. P. Mackenzie}
\affiliation{Scottish Universities Physics Alliance (SUPA), School of Physics and Astronomy, University of St Andrews, North Haugh, St Andrews KY16 9SS, United Kingdom}

\begin{abstract}
We report detailed investigation of quantum oscillations in Sr$_3$Ru$_2$O$_7$, observed inductively (the de Haas-van Alphen effect) and thermally (the magnetocaloric effect). Working at fields from 3~T to 18~T allowed us to straddle the metamagnetic transition region and probe the low- and high-field Fermi liquids. The observed frequencies are strongly field-dependent in the vicinity of the metamagnetic transition, and there is evidence for magnetic breakdown. We also present the results of a comprehensive rotation study. The most surprising result concerns the field dependence of the measured quasiparticle masses. Contrary to conclusions previously drawn by some of us as a result of a study performed with a much poorer signal to noise ratio, none of the five Fermi surface branches for which we have good field-dependent data gives evidence for a strong field dependence of the mass. The implications of these experimental findings are discussed.
\end{abstract}

\maketitle


\section{Introduction}

Quantum criticality is a topic of continued interest in condensed matter physics \cite{ISI000248867000008}. Quantum critical points (QCPs) have been found in various metallic systems, notably heavy fermion intermetallics and transition metal oxides, using tuning parameters such as hydrostatic pressure, chemical doping or magnetic field. They are usually accompanied by a significant enhancement of the low temperature electronic specific heat, which is thought to stem from a quasiparticle mass renormalisation due to strong quantum fluctuations associated with the continuous phase transition\cite{settai}.

QCPs have been studied by a large variety of methods, but surprisingly little is known about the microscopic nature of the processes involved. In particular, it is not clear if, in a multi-band system, the quasiparticles from all the bands will experience the mass renormalisation. The method of choice to determine electronic properties in $k$ space has traditionally been the study of quantum oscillation measurements, which allow, in principle, the determination of the full Fermi surface and the associated quasiparticle masses\cite{shoenberg}, and can be studied over a wide range of magnetic fields. 

The subject of this paper, \TTS, gives an opportunity to probe the nature of quasiparticle mass enhancement in the vicinity of quantum criticality. It features a first order metamagnetic transition which terminates in a critical point\cite{science1}. The depression of its critical temperature towards zero Kelvin using a tilted magnetic field leads to a QCP, situated near 7.9 T when the field is aligned with the crystalline $c$-axis\cite{science1,grigeraPRB}. Anomalous power laws in the resistivity were observed in its vicinity, typical of quantum critical systems \cite{science1}. A pronounced enhancement of the specific heat was reported near the QCP, with a logarithmic divergence of C/T at fields near the critical field \cite{perry1}. Moreover, a previous de Haas-van Alphen (dHvA) study revealed the existence of at least five different electronic bands, of moderately high quasiparticle mass, two of which were reported to become enhanced when approaching the QCP with a magnetic field \cite{borzi}. Away from the critical region, these masses were confirmed in an Angular Resolved Photoemission Spectroscopy (ARPES) study by Tamai $et$ $al.$ \cite{tamai} 

In ultra-pure crystals, the approach to the QCP is signalled by a strong enhancement in the low temperature specific heat \cite{science4} which is cut off by the formation of a new phase, within which transport properties develop a strong anisotropy under the application of tiny in-plane magnetic fields\cite{science3}. In a recent letter we reported observation of dHvA inside this anomalous phase and demonstrated the importance of a chemical potential shift in determining the field dependence of the dHvA frequencies\cite{mercure}.

In this paper, we report a thorough investigation of quantum oscillations in \TTS\ through the critical region. Performing this experiment required the growth of a new generation of crystals, screened in a series of dHvA measurements in order to obtain the highest dHvA signal amplitude. Mean free paths of up to approximately 300~nm were measured, which enabled the detection of the de Haas-van Alphen effect down to fields as low as 2.4 T, as well as the observation of entropy oscillations using the magnetocaloric effect. We present here a detailed study of the quasiparticle masses of five out of six of the known bands as a function of the tuning parameter, the magnetic field. For each of these five bands, \textit{no} significant increase in quasiparticle mass was detected. These results contrast with previous de Haas-van Alphen measurements by some of us, and so we explain the origin of this discrepancy in depth. We furthermore report measurements of the field and angle dependence of the measured dHvA frequencies. The implications of these new results are discussed.

\section{Fermi surface}

\begin{figure}[t]
  \centering
	\includegraphics[width=1\columnwidth]{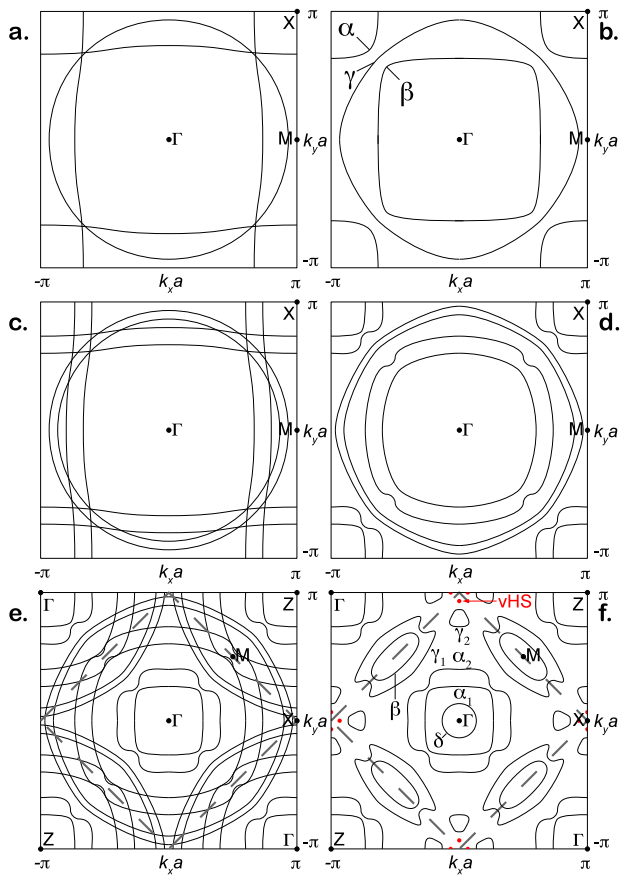}
	\caption{Schematic construction of the Fermi surface of \TOF, $a$ and $b$, and of \TTS, $c$, $d$, $e$ and $f$, where in the last two, the $\sqrt{2} \times \sqrt{2}$ reconstruction of the Brillouin zone (BZ) accompanied by back-folding was included, and the new BZ is represented by the dashed diamond. All axes were scaled by the parameters of the undistorted lattice. The red dots represent the location of a putative van Hove singularity.}
	\label{fig: 327FS2-2}
\end{figure}

The determination of the electronic structure in \TTS\ is not a simple problem, and has not yet been completely solved experimentally. It is a complex system partly due to the high degree of electronic correlations, and partly due to Fermi surface (FS) reconstructions arising from lattice distortions. To introduce the relevant issues, we first show the relationship between the \TTS\ Fermi surface and the simpler one of its single layer relative \TOF. 

\TOF\ possesses a body-centred tetragonal crystal structure, of space group symmetry $I4/mmm$ \cite{mackenzieRMP}. Consequently, it has fourfold rotational and inversion symmetry. Early work on \TTS\ reported it to possess the same space group \cite{ZAAC}, but it was later found using neutron diffraction that this symmetry is broken by a 7$^{\circ}$ rotation of the RuO octahedra \cite{huang,shaked2}. The resulting space group is $Bbcb$, corresponding to a $B$ centred orthorhombic structure with the $a$ and $b$ of sides being effectively $\sqrt{2}$ larger than those of the undistorted crystal. This rotation of the octahedra has a significant effect on the electronic system\cite{ISI:A1997XM97500049}. 

The FS in \TTS\ can be derived schematically without involving band structure calculations, as shown in Fig.~\ref{fig: 327FS2-2}. One starts from the FS in \TOF, where three bands, $\alpha$, $\beta$ and $\gamma$, are present. These originate from the hybridisation of the various $d$ bands of the Ru atoms, the $d_{xz}$, $d_{yz}$ and $d_{xy}$, which, due to crystal field splitting, are the ones crossing the Fermi level.  The $d_{xz}$ and $d_{yz}$ give rise to only weak dispersion either in the $z$ and, respectively, $y$ and $x$ directions, due to quasi one-dimensional hopping. Consequently, the Fermi surface corresponding to these bands should be close to planar in the $k_x k_z$ and $k_y k_z$ directions. These cross in certain regions of the Brillouin zone (BZ), where hybridisation gaps appear, and the sheets reconnect into closed surfaces and produce the $\alpha$ and $\beta$ bands (Fig.~\ref{fig: 327FS2-2}$a.$). The $d_{xy}$ orbital, however,  allows hopping in both the $x$ and $y$ directions, and the corresponding Fermi surface is close to a perfect circle in the $k_x k_y$ plane, the $\gamma$ band. The resulting FS with all bands is shown in Fig. \ref{fig: 327FS2-2}$b$. This construction is consistent with the fourfold rotation symmetry of the $I4/mmm$ space group.

In \TTS, one expects each band to duplicate, due to bilayer splitting \footnote{After distortion, \TTS\ possesses four instead of two non-equivalent Ru atoms, bringing the number of $d$ bands crossing the Fermi level from three to six.}. Consequently, the result is slightly different than for \TOF. One starts from six surfaces (Fig.~\ref{fig: 327FS2-2}$c.$), four that originate from the $d_{xz}$, $d_{yz}$ orbitals, which reconnect at the points of crossing, and one should obtain a result similar to that shown in Fig.~\ref{fig: 327FS2-2}$d$. This is what the FS would be without the $\sqrt{2} \times \sqrt{2}$ reconstruction due to the octahedral rotation. But since the BZ reconstructs into a square of half the area, back-folding of the bands occurs, shown in Fig.~\ref{fig: 327FS2-2}$e$. In this plot, many band crossings appear, and the way in which the surfaces reconnect is complex. According to dHvA and ARPES experiments\cite{borzi, tamai}, six orbits are expected, shown in Fig. \ref{fig: 327FS2-2}$f$. These take the form of square and cross shaped hole pockets in the centre, originating from the $d_{xz}$ and $d_{yz}$ orbitals, two lens shaped electron pockets at the $M$ point, resulting from hybridisation between $d_{xy}$ and $d_{xz}$, $d_{yz}$, and a small pocket near the $X$ point. These were labelled, respectively, $\alpha_1$, $\alpha_2$, $\gamma_1$, $\beta$ and $\gamma_2$ (see Fig.~\ref{fig: 327FS2-2}$f$). Finally, there is a small circular pocket centred on $\Gamma$, labelled $\delta$, that has no direct analogy in \TOF.  It has strong Ru $d_{x^2-y^2}$ character, and results from the depression in energy of a band that is unfilled in \TOF\ due to the lower symmetry of \TTS\ \cite{ISI:A1997XM97500049}.

\section{Experiment}

\TTS\ crystals were produced using a floating zone image furnace, using the method published by Perry and Maeno\cite{perryGrowth}. 10 crystal rods were grown, from which more than 50 individual samples were cut. Each was nearly cylindrical, 2~mm in diameter and 3~mm long, with the crystalline $c$-axis along the principal axis. These were the subject of a series of characterisation experiments, which enabled us to select the best four to use in these experiments, according to two criteria: disorder, due to vacancies and impurities, and phase purity, affected by the appearance of intergrowth regions with $n \neq 2$ consecutive RuO planes (\OOT, \TOF, \FTT, etc.). 

Low disorder crystals were selected in a two-step process in which preliminary residual resistivity measurements were followed by short dHvA experiments performed in an adiabatic demagnetisation refrigerator. This featured the use of first harmonic detection, a DC field between 6 and 7~T and a temperature of 250~mK. The samples selected for the main experiments were those showing the largest amplitude of the 1.78~kT frequency (see Table \ref{tab: FreqMasses}).


In order to determine impurity phase volume fractions, we used the magnetic properties of the competing phases: superconductivity below 1.5~K ($n = 1$, \TOF )\cite{mackenzieRMP}, ferromagnetism below 160~K ($n = \infty$, \OOT )\cite{Kanbayasi, Allen, Cao2}, ferromagnetism below 105~K ($n = 3$, \FTT )\cite{Cao}. Since \TTS\ is paramagnetic\cite{Ikeda2000} and produces only a small signal in susceptibility and magnetisation experiments, these properties were easily detected quantitatively for small amounts using zero field AC susceptibility (superconductivity) and magnetisation measurements using a Quantum Design MPMS SQUID magnetometer (ferromagnetism). The samples used in the quantum oscillation experiments featured residual resistivities typically of less than 0.4~$\mu \Omega$cm, and impurity phase volume fractions of a few \% or less for \TOF\ and 0.1\% for \OOT\ and \FTT.

\begin{figure}[t]
  \centering
	\includegraphics[width=1\columnwidth]{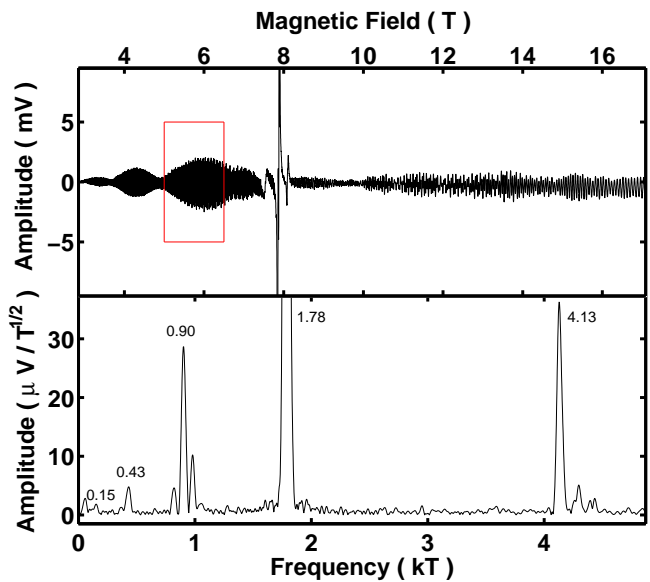}
	\caption{$Top$ $panel$ Typical second harmonic dHvA oscillations measured with B$||c$-axis, at a temperature of 55~mK (see Mercure $et$ $al.$ for a discussion of the thermometry in this system\cite{mercure}). $Bottom$ $panel$ Fourier transform of the data in the top panel between 5 and 6.5~T (shown as a red square). Five frequencies were observed, at 0.15, 0.43, 0.90, 1.78 and 4.13~kT.}
	\label{fig: Spectrum1}
\end{figure}

The de Haas-van Alphen effect was measured using a dilution refrigerator located in the Cavendish Laboratory in Cambridge, equipped with an 18~T superconducting magnet. The samples were placed in well compensated coil pairs, each comprising 1000 turns of 25$\mu$m diameter Cu wire. These were placed onto a rotation mechanism which allowed the measurement of the magnetic susceptibility at angles between the field and the crystalline $c$-axis of up to about 60$^{\circ}$, both along the [100] and [110] directions. The coil pairs were connected to the input coils of low temperature transformers, which were themselves coupled with room temperature preamplifiers. The magnetic susceptibility was measured using first and second harmonic detection, with digital lock-in amplifiers and an excitation field of 20 G at 7.9 Hz. These values for the modulation field were chosen in order to ensure that eddy current heating in the samples was negligible. They were selected following an extensive study which demonstrated both that the measured masses were independent of modulation fields up to this level, and that, under the conditions selected for the experiments on \TTS, the well-known masses of all sheets of \TOF\ were obtained correctly. The noise floor at the compensated coils was of around 30~pVrms/$\sqrt{\mathrm{Hz}}$, when using an analog magnet discharge field sweep method, and 150~pVrms/$\sqrt{\mathrm{Hz}}$ when using linear field sweeps. 

For the majority of the measurements presented here, second harmonic detection was employed in order to minimise the effects of the large magnetic background near the metamagnetic transition. A series of sweeps was performed between 18~T and 2.5~T, at rates of approximately 0.04~T/min for B$>$10~T and 0.02~T/min for lower fields. These were carried out at 10 different temperatures between 90 and 500~mK with the field aligned with the crystalline $c$-axis, and then at angles from -15 to 55$^{\circ}$ using a step of about 1.6$^{\circ}$, at a temperature of approximately 55~mK. For one aspect of the data set, the field dependence of very low dHvA frequencies at low fields, we performed additional measurements. Oscillations were studied in St Andrews using the magnetocaloric effect in which the sample temperature is monitored while the field is swept. The relaxation times associated with this kind of measurement favour low frequencies over higher ones. To follow those up with dHvA we worked with first harmonic detection, maximising the low frequency signal at the expense of a large magnetic background which restricted the experiment to fields below 7 T.

\section{dHvA data, spectra and quasiparticle masses}

In Fig.~\ref{fig: Spectrum1} we present susceptibility data between 3~T and 18~T and the dHvA spectrum of data between 5~T and 6.5~T. The top panel shows that dHvA oscillations can be detected throughout the available field range. Three metamagnetic signatures can be seen near 8~T, which consist of asymmetric peaks centred at 7.5, 7.9 and 8.1~T. The first corresponds to a metamagnetic cross-over and the second to two first order transitions, as determined using the out-of-phase component of the first harmonic AC susceptibility (not shown)\cite{grigeraPRB}. The region between 7.9 and 8.1~T corresponds to the anomalous phase discussed by Borzi $et$ $al.$ \cite{science3} and our recent work in which we reported dHvA oscillations\cite{mercure, MercureThesis}. The dHvA spectrum shown in the lower panel of Fig.~\ref{fig: Spectrum1} features five peaks, situated at 0.15, 0.43, 0.90, 1.78 and 4.13~kT. In previous ARPES work by some of us\cite{tamai}, we associated these with bands labelled respectively $\beta$, $\delta$, $\gamma_1$, $\alpha_1$ and $\alpha_2$, where the first three correspond to electron orbits and the last two to hole orbits. A sixth frequency also exists at 110~T, as will be discussed in sec. \ref{sec:magcalosc}. Note that the splitting of the peak at 0.9 kT into four peaks is associated with bilayer splitting of this band, as well as to a normal corrugation of this branch of the FS, while the splitting of the peak at 4.13~kT is due to magnetic breakdown, as detailed in section VI.

Taking into consideration all the bands reported, the ARPES map of the BZ and the predicted multiplicity and bilayer splitting of these bands, we were able to put forward a model which allowed us to perform the Luttinger sum and account for the total electronic specific heat coefficient in \TTS, which was reported by Ikeda $et$ $al.$ as 110 mJ/molRuK$^2$\cite{Ikeda2000}. It involves assigning multiplicities to bands and determining whether they are electron or hole-like. Table \ref{tab: FreqMasses} shows a summary of this model, along with the dHvA frequencies of the six bands, the associated number of electrons which occupy these, their associated quasiparticle masses and their contribution to the total specific heat. The number of electrons was calculated by normalising the dHvA frequencies by that corresponding to the in-plane area of the BZ, 13.7~kT, multiplied by the multiplicity of the bands, and taking into account the spins. For hole-like bands, these are expressed as the number of filled zones minus the fraction of electrons covered by these bands. The contributions to the specific heat were calculated using the quasiparticle masses, times the multiplicity, and a conversion factor of 0.74 mJ/molRuK$^2$\cite{PhysRevLett.76.3786}. Moreover, a number of electrons are included in zones that become filled in the process of the back-folding of the bands, which accounts for 8 electrons. The total number of electrons counted in this way nearly matches the required amount, 15.71 compared to 16 electrons, while the calculated specific heat is consistent with that measured previously, $103\pm8$ compared to 110~mJ/molRuK$^2$.

\begin{table}[t]
\begin{center}
		\begin{tabular*}{1\columnwidth}{@{\extracolsep{\fill}}c c c c c c c}
			\\
			\hline
			Label & Type & Mult. & $F$ & Electrons & $m^*/m_e$ & $\gamma_{el.}$\\
			  &   &   &  (kT)  &   &  & mJ/molRuK$^2$\\
			\hline
			$\gamma_2$ & $h$ & 8 & 0.11 & 4-0.128 & $8 \pm 1$ & $47\pm6$\\
			$\beta$ & $e$ & 2 & 0.15 & 0.044 & $5.6 \pm 0.3$ & $8.3\pm0.4$\\
			$\delta$ & $e$ & 2 & 0.43 & 0.126 & $8.4\pm0.7$ & $12\pm1$\\
			$\gamma_1$ & $e$ & 4 & 0.91 & 0.532 & $7.7\pm0.3$ & $23\pm1$\\
			$\alpha_1$ & $h$ & 1 & 1.78 & 2-0.260 & $6.9\pm0.1$ & $5.1\pm0.1$\\
			$\alpha_2$ & $h$ & 1 & 4.13 & 2-0.602 & $10.1\pm0.1$ & $7.5\pm0.1$\\ 
			$Filled$ & $e$ & 8 & 13.7 & 8 &&\\
			$Total$ &&&& 15.71 &  & $103\pm8$\\
			\hline
		\end{tabular*}
	\caption{Electronic data for all six bands observed in \TTS. The type refers to whether bands are electron-like ($e$) or hole-like ($h$) and the multiplicity refers to possible bilayer splitting times its number of occurrences inside the BZ. $F$ is the dHvA frequency, the number of electrons refers to the fraction of the BZ area covered by a band, which includes the multiplicity. Finally, $m^*/m_e$ is the quasiparticle mass and $\gamma_{el.}$ is the associated contribution to the electronic specific heat coefficient, including the multiplicity.}
	\label{tab: FreqMasses}
\end{center}
\end{table}

\section{Magnetothermal detection of low field, high mass oscillations}
\label{sec:magcalosc}
\begin{figure}[t]
  \centering
	\includegraphics[width=1\columnwidth]{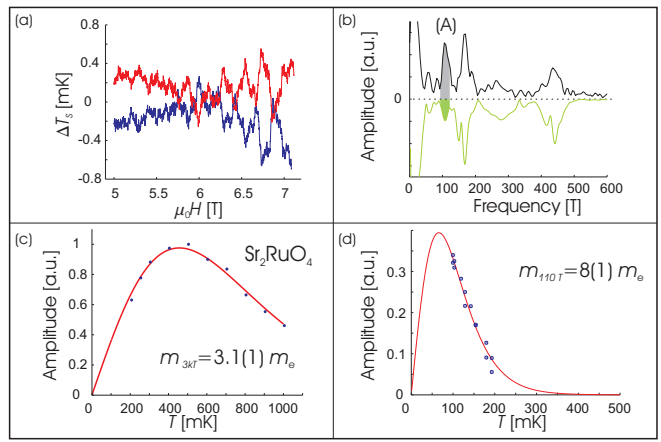}
	\caption{\textit{(a)} The change of sample temperature $T_S$ during a sweep of the magnetic field at 0.04~T/min (blue) and -0.04~T/min (red) as a function of magnetic field $\mu_0H$. The average sample temperature is 160~mK. \textit{(b)} Fourier transform in
1/$\mu_0H$ of the blue trace in panel (a). In green is given the Fourier transform of a dHvA measurement over a comparable magnetic field region at base temperature. The label (A) indicates the new frequency found at 110~T. \textit{(c)} Temperature dependence of the amplitude of the 3~kT frequency of \TOF\ as measured by magnetothermal oscillations. The red curve is a fit of the temperature
derivative of the Lifshitz-Kosevich formula to the data. \textit{(d)} Amplitude of the 110~T peak in the Fourier transform of the data in panel (a) as a function of sample temperature $T_S$. The red curve is a fit of the temperature derivative of the Lifshitz-Kosevich function to the data. This fit gives a mass $m_{110T}$ for the newly observed frequency of 8(1)~$m_e$.}
	\label{fig: magnetooscillations}
\end{figure}

One of the predictions of the ARPES work of Tamai et al. \cite{tamai} was the possible existence of tiny Fermi surface sheets associated with extremely flat bands that they labelled $\gamma_2$. Observing tiny frequencies against the background of many other oscillatory components is not easy, especially since the flatness of the bands means that for any given field range, the signal size is likely to be suppressed relative to that of other components of the overall signal. This effect is exacerbated when working with second harmonic dHvA, in which relative signals for low frequencies go as $F^2$.  In order to search for the oscillations, we profited from bespoke equipment developed in St Andrews for the study of the magnetocaloric effect \cite{science4}. In contrast to second harmonic dHvA, the relative signal size will be independent of $F$ in principle, but subject to a damping factor that will grow rapidly with $F$ since more rapid oscillations will be damped more strongly in a measurement system with long intrinsic relaxation times.

In Fig. \ref{fig: magnetooscillations}a we show low frequency oscillations detected by the magnetocaloric effect for up- and down-sweeps of the field in red and blue respectively.  The sign reversal of the temperature change is a key diagnostic that these are genuine magnetothermal oscillations. If, for example, they were due to sample heating due to eddy currents subject to the Shubnikov-de Haas effect, they would have a sign independent of field sweep direction.  Their Fourier transformation is shown in the black trace in Fig. \ref{fig: magnetooscillations}b.  The previously undiscovered peak at 110~T is shaded.  In the same panel, the green trace shows the Fourier spectrum of a subsequent dHvA run taking data on the first harmonic with a modulation field of 3~gauss.  While confirming the existence of the 110~T peak, it emphasises the relative signal enhancement that magnetocaloric oscillations can provide for very low frequency signals. 

An important aspect of magnetocaloric oscillations is that the amplitude of the signal is proportional to $dM/dT$ rather than to $M$ directly \cite{shoenberg}, so the temperature dependence expected of the oscillations is that of the first thermal derivative of the usual Lifshitz-Kosevich form (see equation \ref{eq:LK} below).  This is shown in red in Fig. \ref{fig: magnetooscillations}c which shows the result of a 'control' experiment detecting magnetocaloric oscillations from the well-known $\alpha$ sheet of \TOF. In spite of the different temperature dependence of the raw signal, the mass deduced agrees well with that obtained by standard dHvA, providing a useful verification of the calibration of the magnetocaloric apparatus. 

In Fig. \ref{fig: magnetooscillations}d we show the temperature dependence of the magnetocaloric signal of the new frequency at 110~T, along with the fit that allows an effective mass estimate of 8$\pm$1~m$_e$. A subsequent first harmonic dHvA study involving careful signal averaging increased the precision of the measurement to 7.2$\pm$0.2$m_e$. This mass corresponds to a Fermi velocity of only 9~kms$^{-1}$, a factor of six smaller than that of the $\gamma$ sheet of \TOF.

\section{Field dependence of dHvA frequencies and possible magnetic breakdown}

\begin{figure}[t]
  \centering
	\includegraphics[width=1\columnwidth]{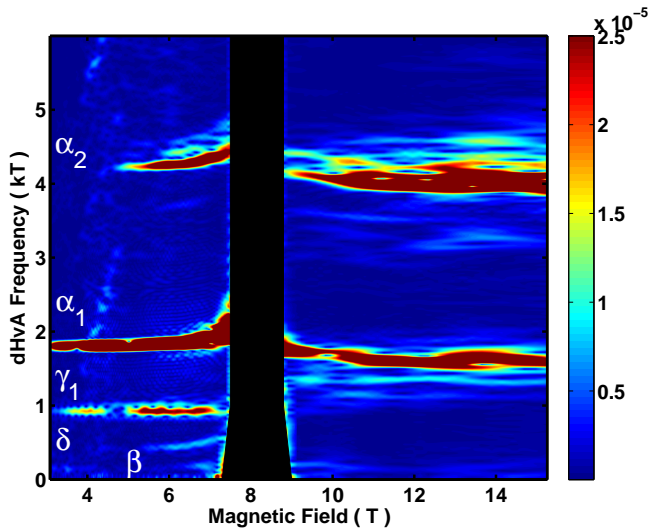}
	\caption{Field dependence of the dHvA spectra observed with the field aligned with the direction of the $c$-axis. The colour scale corresponds to the amplitude of the spectrum.}
	\label{fig: FvsB}
\end{figure}

As was already suggested by Borzi $et$ $al.$\cite{borzi}, and described in more detail in our recent work\cite{mercure}, the dHvA frequencies are not constant along the field range, but change as the system crosses the metamagnetic transition. Variations are observed as one increases the magnetic field and are especially significant above 7~T. 

In Fig.~\ref{fig: FvsB} we present an analysis of the field dependence of the dHvA frequencies along the field range. It was calculated using multiple Fourier transforms taken over short field ranges of equal inverse field width. 100 such windows were used, of width of $0.02$~T$^{-1}$, which overlap and are distributed evenly in inverse field space\footnote{The inverse field width of each window is about 3.6 times larger than the inverse field spacing between the centres of consecutive windows.}. The intensity of each spectrum calculated is plotted on a colour scale. The hidden central area corresponds to the metamagnetic region, where a large non-oscillatory signal arises (see Fig.~\ref{fig: Spectrum1} near 8~T). We have recently succeeded in detecting a tiny oscillatory signal on top of the large background, as reported in ref. [~\onlinecite{mercure}].

On the low field side of the metamagnetic transition, the same five dHvA frequencies are observed as in Fig.~\ref{fig: Spectrum1} (excluding $\gamma_2$). However, near 7~T, increases are observed, particularly with the 1.78 and 4.13~kT frequencies, which correspond to the $\alpha_1$ and $\alpha_2$ bands. This is discussed in detail in ref. [~\onlinecite{mercure}], and corresponds to a change in the size of these parts of the FS, due to a change in Fermi level. Note that the fact that the $\alpha_1$ and $\alpha_2$ bands remain of similar size in high fields compared to low fields exclude any further $\sqrt{2} \times \sqrt{2}$ reconstruction of the FS at the metamagnetic transtion. 

One can also observe in Fig.~\ref{fig: FvsB} that the data become more complex as the magnetic field increases, and that most dHvA peaks split into several components. The new frequencies might originate from magnetic breakdown, activated as the field increases. In particular, for the case of $\alpha_2$ (see Fig.~\ref{fig: 327FS2-2}$f$.), symmetry considerations allow four magnetic breakdown regions with the $\gamma_1$ pocket, which increase the cross-sectional area by small amounts. This leads to a large number of possibilities, each new orbit occurring with a slightly different probability. In such a case, the amplitude of the dHvA signal becomes divided between a large number of possible frequencies all situated near 4 kT. The same reasoning holds with the $\alpha_1$ pocket, from which the electrons can tunnel towards the $\alpha_2$ pocket and back in four different locations, leading to a similar number of possibilities. 


It is difficult to explore this in any more detail, since the topology of the Fermi surface obviously changes slightly across the metamagnetic transition and is not known above 8~T. It is thus impossible to know the magnetic breakdown field values precisely. We note however that, as can be seen in Fig.~\ref{fig: FvsB}, the frequency splitting that we associate with magnetic breakdown appears at fields lower than that of the metamagnetic transition, with the 4.1~kT peak splitting into three or more between 6~T and the metamagnetic transition. This suggests that the breakdown and the metamagnetism are not directly related\footnote{One could also invoke quantum interference oscillations as a source of splitting of dHvA peaks. However, as shown in \cite{shoenberg}, these are by nature insensitive to temperature. We can show with high temperature data (not shown in the paper) that no oscillations remain above 800 mK over the whole field range from 18 to 2 T.}.

\section{Angle dependence of dHvA frequencies}

\begin{figure}[t]
  \centering
	\includegraphics[width=1\columnwidth]{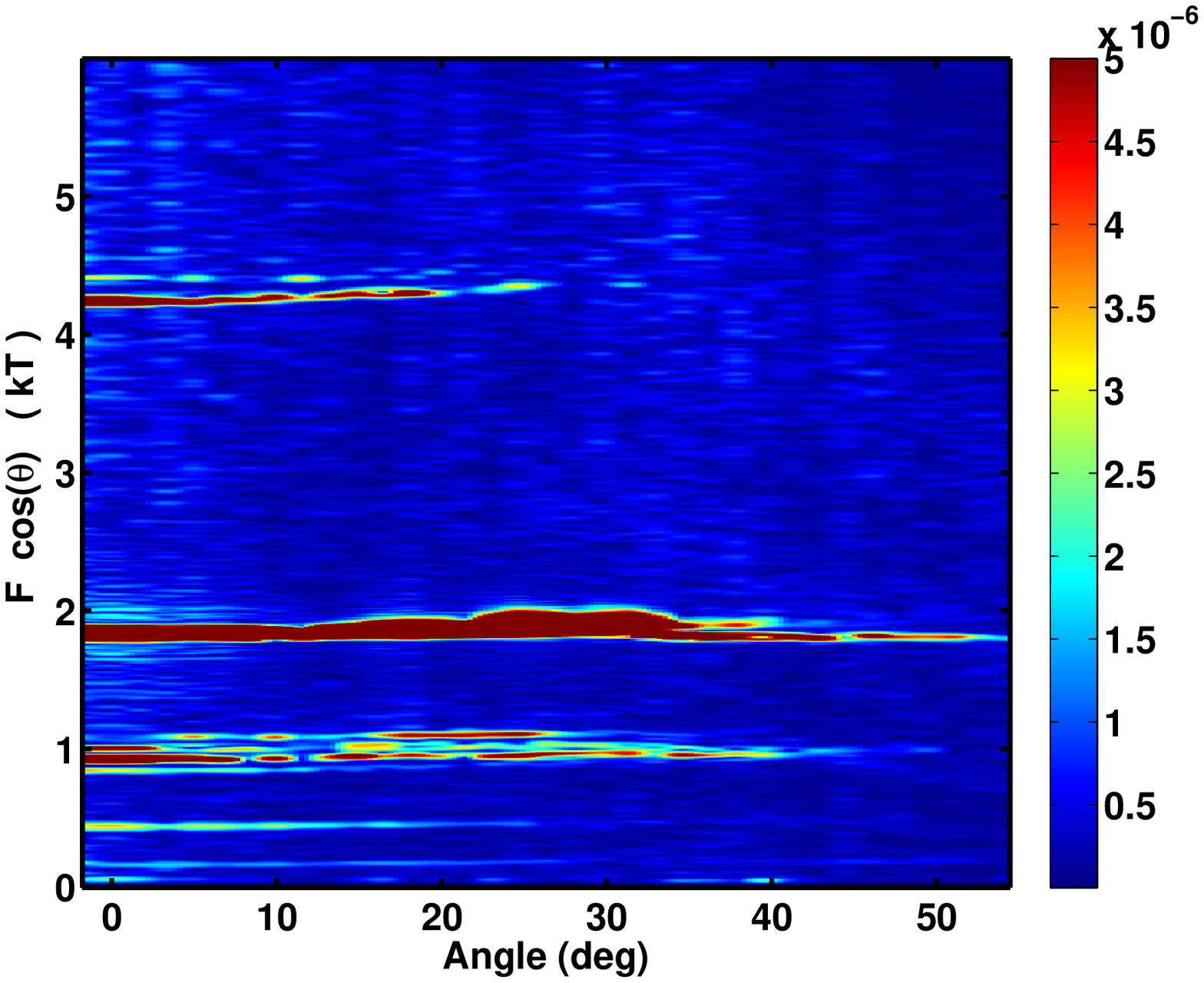}
	\includegraphics[width=1\columnwidth]{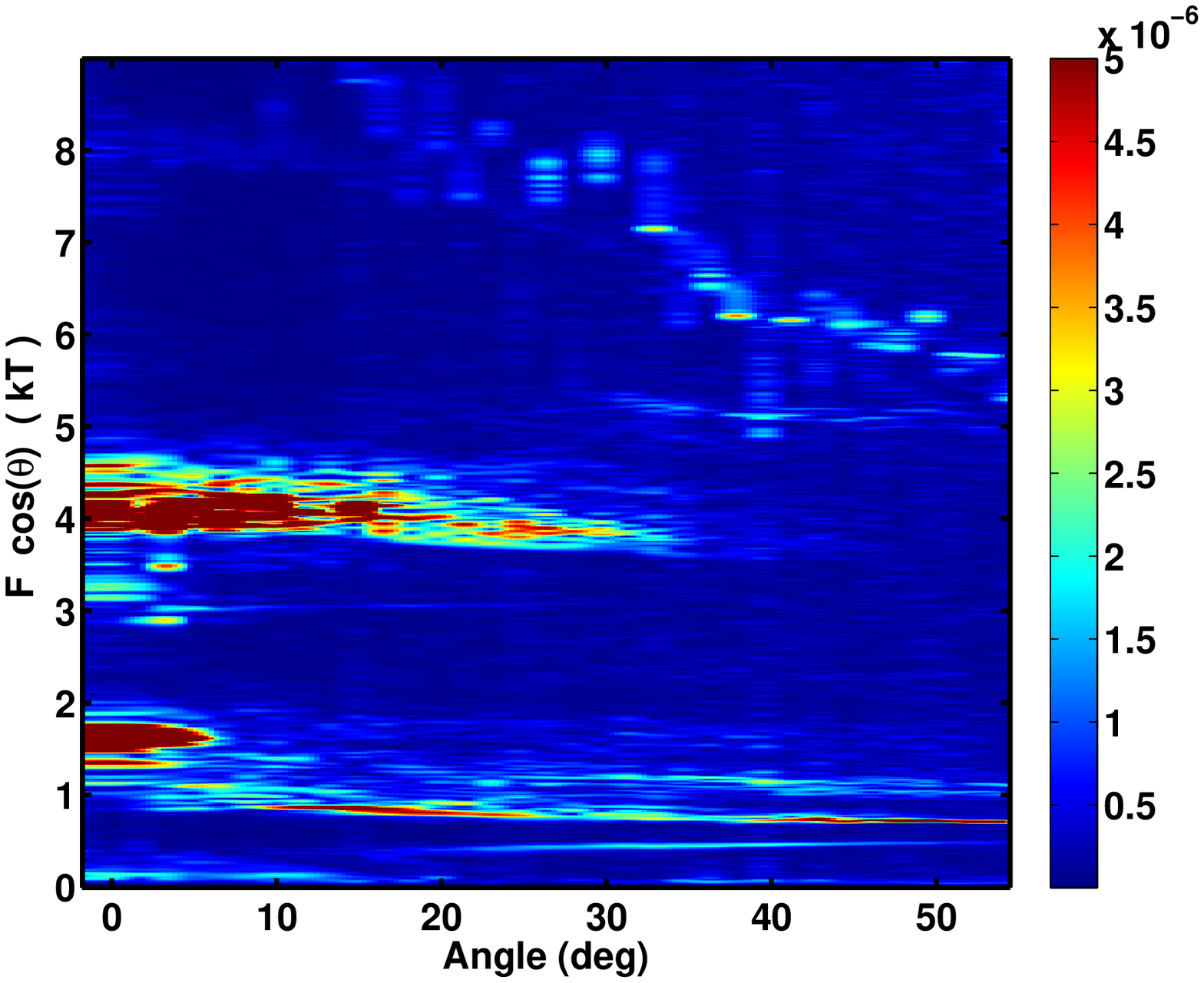}
	\caption{\textit{Top panel} Angular dependence of the dHvA spectra observed in the low field side of the metamagnetic transition, where the colour represents the spectral intensity, while the horizontal and vertical axes represent the angle and the dHvA frequency multiplied by $\cos(\theta)$ respectively. Fourier transforms were taken between 5 and 6.5~T. \textit{Bottom panel} Angular dependence of the spectra in the high field side of the metamagnetic transition. Fourier transforms were taken between 10 and 18~T. }
	\label{fig: Angular}
\end{figure}

Measurements of the angle dependence of the dHvA signal allow one, in principle, to determine the three dimensional topology of the FS. In quasi-two dimensional materials, orbits are open in the $k_z$ direction, and the frequencies follow a simple angular dependence of the form $F(\theta) = F_0/ \cos \theta$, where $F_0$ is the frequency when the field is aligned with the $c$-axis direction\cite{bergemann}. This is observed in the case of \TTS, for fields lower than that of the metamagnetic transition. Fig.~\ref{fig: Angular}, top panel, presents Fourier transforms of the data between 5 and 6.5~T measured at angles between 0 and 54$^{\circ}$, plotted with the frequency multiplied by $\cos(\theta)$, such that cylindrical FS pockets appear as horizontal lines. The amplitude of the spectra are represented in colour in vertical strips for each angle. All of the FS pockets observed follow almost perfectly the expected behaviour for a 2D system, and small deviations are due to their small warping\footnote{Note that the data excludes orbits closed near the top of the zone, which would show up, for instance, in the data of the 1.79 kT peak ($\alpha_1$), since the orbits reach the top of the zone at 52$^{\circ}$ (the height of zone is $k_z$ = 600 A$^{-1}$, while the in-plane Fermi vector is $k_F$ = 234 A$^{-1}$, thus one reaches the top of the zone at $\tan^{-1}(300/234)$ = 52$^{\circ}$), an angle covered by our data. The $\alpha_1$ orbit, if closed near the top of the zone, would show a downturn of the frequency at an angle below 52$^{\circ}$. Oscillations were searched at higher angles than those shown in this paper, but none were detected.}. This agrees well with the predictions made by band structure calculations \cite{tamai,singh}.

The situation is not quite the same at fields above that of the metamagnetic transition. We found that frequencies are not continuous functions of angle; some of them appear while others vanish at specific angles. In particular, the frequency at 1.6~kT, related with the $\alpha_1$ pocket, disappears between 5 and 10$^{\circ}$, while that at 1.0~kT, probably related to the $\gamma_1$ pocket, is not present for $B\left\|c\right.$-axis and appears at a similar angle. We observed a similar behaviour with the frequency near 0.43~kT, which appears near 20$^{\circ}$. Finally, frequencies near 3~kT appear briefly at low angles. These phenomena may also be related to the magnetic breakdown effects mentioned earlier. Effectively, when a two dimensional FS possesses small corrugations, magnetic breakdown fields can vary as the field angle is changed, due to a changing $k$-space distance between FS pockets. If such a distance decreases significantly, magnetic breakdown becomes overwhelmingly probable, to the extent that the original orbit becomes less probable, and may escape detection. Varying magnetic breakdown gaps might therefore make the angle dependence of dHvA frequencies appear as if the FS changed topology with angle, while no such changes occur in reality.

\section{Quasiparticle masses}

\begin{figure}[t]
  \centering
	\includegraphics[width=1\columnwidth]{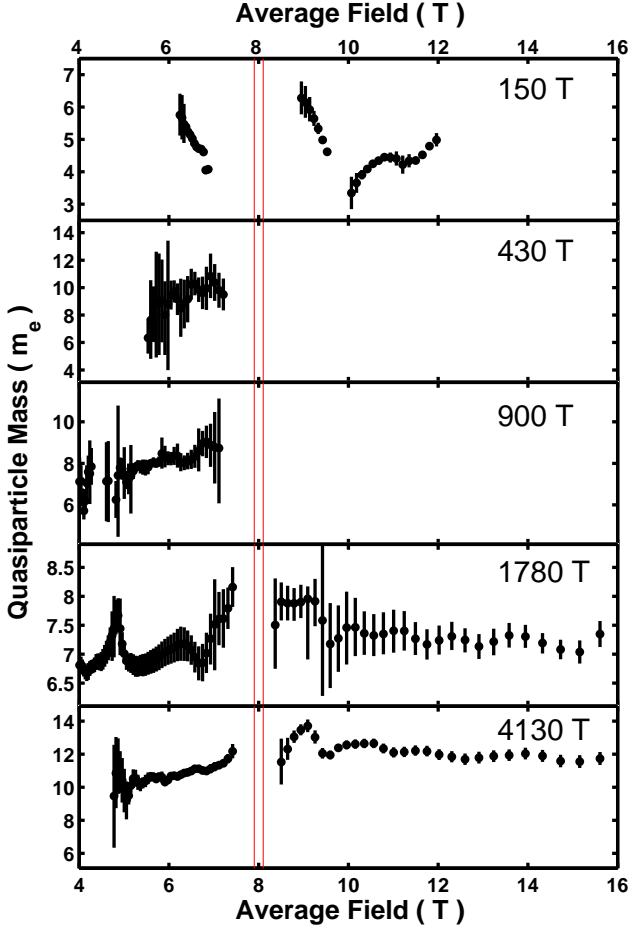}
	\caption{Quasiparticle mass of the five frequencies identified in the low field side, as function of applied magnetic field. These correspond respectively to the bands $\alpha_1$, $\alpha_2$, $\delta$, $\gamma_1$ and $\beta$. The anomalous phase region, situated between 7.9 and 8.1~T, is indicated with vertical red lines. The fits were calculated over field ranges of 0.01~T$^{-1}$ except for the 0.15~kT frequency, for which a width of 0.02~T$^{-1}$ was necessary in order to properly resolve the corresponding peaks in the spectra. Gaps in the data are due to the amplitudes becoming too small to fit because of beat patterns in the data.}
	\label{fig: FirstMassA}
\end{figure}

\begin{figure}[t]
  \centering
	\includegraphics[width=.98\columnwidth]{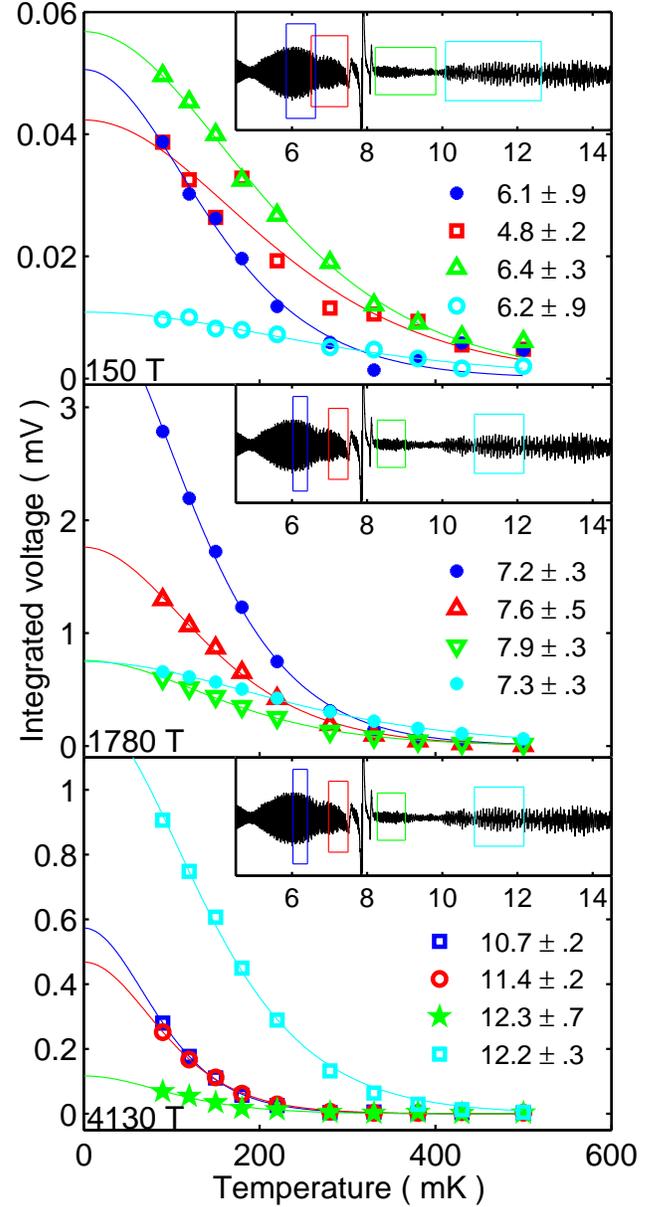}
	\caption{LK fits performed in various field ranges for three of the six detected frequencies, 0.15, 1.78 and 4.13~kT, which correspond to the $\beta$, $\alpha_1$ and $\alpha_2$ bands. As in the data of Fig. \ref{fig: FirstMassA}, for the case of 0.15~kT, inverse field windows of a width of 0.02~T$^{-1}$ were used, while for the other two the width was of 0.01~T$^{-1}$. The field regions are indicated in colour boxes in the respective insets, and the same colour schemes are respected for the data. Integrated peaks in the Fourier spectra are represented with various symbols, while fitted LK curves are shown with solid lines. The values in the legends correspond to the extracted quasiparticle masses.}
	\label{fig: LKFits1}
\end{figure}


The enhancement of the specific heat previously reported near the metamagnetic transition is expected, in a metal, to be reflected in the quasiparticle masses measured from quantum oscillations \cite{perry1}. Borzi $et$ $al.$ reported strong enhancements of the quasiparticle mass of the $\alpha_1$ and $\alpha_2$ bands, below and above the metamagnetic transition\cite{borzi}. It was moreover expected that, although not possible to determine at the time, the masses of the other bands should be enhanced as well, stemming from quantum critical fluctuations coupling to the whole Fermi sea. Indeed, the desire to confirm and extend the findings of ref. [~\onlinecite{borzi}] was the main factor motivating this new study using samples and detection methods designed to deliver an improved signal to noise ratio. To our surprise, we did not reproduce our previous findings. In contrast, we demonstrate here that no such enhancement occurs for the five bands which we detected in second harmonic dHvA; $\alpha_1$, $\alpha_2$, $\delta$, $\gamma_1$ and $\beta$ (see Fig.~\ref{fig: 327FS2-2}$f$). We furthermore show that the data analysis method used by us in ref.~[\onlinecite{borzi}] to calculate quasiparticle masses can lead to erroneous results if used with data of low signal to noise ratio, as we discuss in detail in appendix \ref{App:B}. 

In Fig. \ref{fig: FirstMassA} we present the field dependence of the quasiparticle masses of the dHvA frequencies at, in order from top to bottom, 0.15, 0.43, 0.90, 1.78 and 4.13~kT, which correspond, respectively, to $\beta$, $\delta$, $\gamma_1$, $\alpha_1$ and $\alpha_2$ (excluding $\gamma_2$), using a magnetic field aligned with the crystalline $c$-axis. The field dependent masses were calculated using 100 field windows of equal inverse field width of 0.01~T$^{-1}$ between 4 and 18~T, over which Fourier transforms were performed for all measured temperatures. The amplitude of the oscillations at specific frequencies was obtained by integrating peaks in the power spectra. These were subsequently analysed using the standard Lifshitz-Kosevich (LK) relation
\beq
LK(B_0,T) = A{Cm^*T/B_0 \over \sinh(Cm^*T/B_0)}, \quad B_0 = {2\over 1/B_1 + 1/B_2},
\label{eq:LK}
\eeq
where $A$ is a temperature independent factor, $B_0$ is the inverse of the average inverse magnetic field, $C$ contains all universal constants and is approximately equal to 14.7~T/K$m_e$ and $m^*$ is the quasiparticle mass. Note that the change to $B_0$ due to the increase in internal field around the metamagnetic transition is of less than 0.2\% and was neglected. Non-linear fits of this relation to the data were performed using two parameters, $A$ and $m^*$, for each field window, and each fit was carefully inspected by eye, rejecting field windows which produced inappropriate data distributions. The resulting masses are plotted in Fig.~\ref{fig: FirstMassA} as a function of $B_0$. The error bars were obtained from the standard non-linear regression procedure. Fig.~\ref{fig: LKFits1} shows four of these LK fits for the frequencies at 0.15, 1.78 and 4.13~kT in selected field regions, which are shown as colour boxes in the respective insets. The fits are mostly very good.

It is clear from the data in Fig.~\ref{fig: FirstMassA} and from the fits in Fig.~\ref{fig: LKFits1} that no strong enhancements of the quasiparticle masses of the bands $\beta$, $\delta$, $\gamma_1$, $\alpha_1$ and $\alpha_2$ are present near the metamagnetic transition. Changes of at most 4 electron masses were seen, which originate from fitting errors when the signal to noise decreases at, for instance, nodes of beating patterns. For the case of $\gamma_2$, situated at 0.11~kT, its resolution from the peak at 0.15~kT required a field range on the low field side of around 4 to 7~T, making the calculation of the field dependence of its mass impossible. The second harmonic data shown here suppress its signal so much in relation to that of the 0.15~kT frequency that the temperature dependence of the latter could be successfully isolated. Note furthermore that the absence of data for the peaks at 0.43 and 0.90~kT on the high field side of the metamagnetic transition is due to the absence of signal in that region for these bands (see Fig.~\ref{fig: FvsB}). The signal for these bands appears at higher angles (see Fig.~\ref{fig: Angular}), suggesting that another process suppresses the amplitude of their quantum oscillations at low angle but that these bands are still present in that region. This result has been reproduced several times, and additional verifications have been performed using first harmonic detection and a lower modulation field (3~G compared to 20~G).

The absence of a quasiparticle mass enhancement as a function of magnetic field in \TTS\ is a surprise to us. Both the electronic specific heat \cite{science4} and the weight of the Fermi liquid $T^2$ term \cite{science1} show approximately consistent enhancements  that, together, imply an average rise of the quasiparticle mass of a factor of between 2 and 3 between low fields and 7.5~T.  It was therefore vital that we checked the sensitivity of our analysis techniques to an enhancement of this scale for all the detectable frequencies in our signal. We decided that the only way to test this in practice was using simulated dHvA data. Starting with the same parameters as those of our experiment (including modulation field, data sampling rate, relative amplitude factors for the different oscillatory components and estimated sample mean free path), we modelled a composite signal comprising the $\beta$, $\delta$, $\gamma_1$, $\alpha_1$ and $\alpha_2$ frequencies to which we added random uncorrelated Gaussian noise of a similar magnitude to that present in our experiments.

\begin{figure}[t]
  \centering
	\includegraphics[width=1\columnwidth]{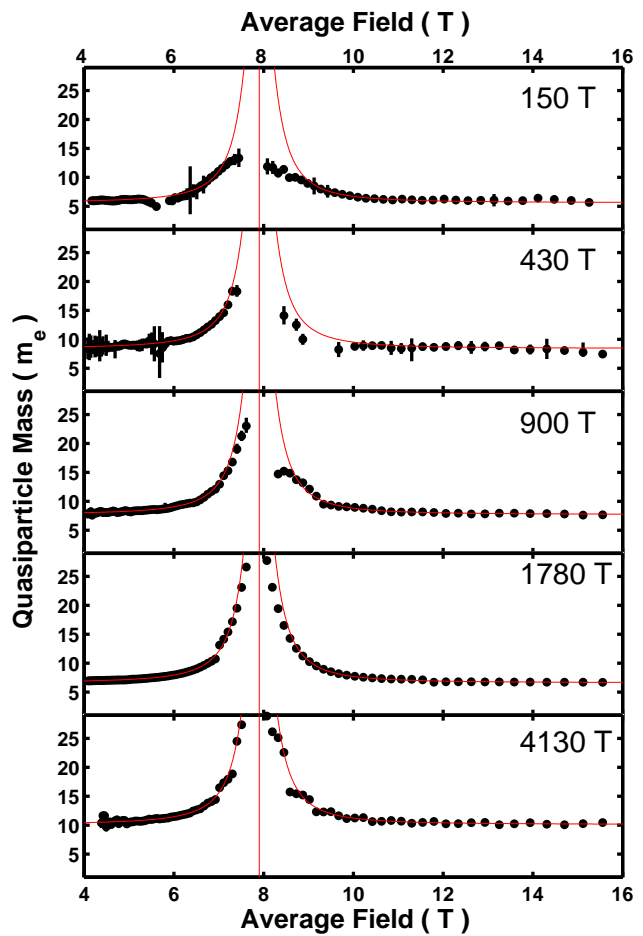}
	\caption{Quasiparticle mass of five simulated dHvA frequencies similar to those detected in our experiments, calculated using the same method as that used in Fig.~\ref{fig: FirstMassA}. The data was modelled using quasiparticle masses which were strongly enhanced near 8~T. The oscillations were calculated using the masses and mean free path extracted from the experiment, as well as amplitudes, field dependent frequencies and beat patterns similar to those in the data and uncorrelated gaussian distributed noise was added to the oscillations.}
	\label{fig: MassesDiv2}
\end{figure}

The simulated data were subsequently analysed using exactly the same procedure as for the experimental data. The first data set used a field varying quasiparticle mass for the five bands simulated, using the experimental value away from the metamagnetic transition, and a strong, gradual enhancement shown by the red lines in Fig.~\ref{fig: MassesDiv2}. The onset of this increase was determined approximately from that of specific heat data \cite{science4}. In contrast, the second simulation featured no quasiparticle mass enhancements, but instead an increase in scattering rate of a factor of 2 at 8~T in order to replicate the observed decrease in signal at the metamagnetic transition (see Fig.~\ref{fig: Spectrum1}, top panel). 

In Fig.~\ref{fig: MassesDiv2} we present the results of the analysis of the simulated data featuring mass enhancements for all the bands. The red curves indicate the values of the mass used in the calculations. We observed that depending on the signal to noise of the various oscillatory components, the extracted masses track the original ones very well for mass changes of up to around 10~$m_e$ for the component with lowest signal to noise ratio (0.15~kT), to above 25~$m_e$ for that with the highest quality signal (1.78~kT), all of which are much larger than the largest mass changes measured in our data (of at most 4~$m_e$). We thus conclude that if such mass changes had been present in our dHvA data, we would have detected them. 

\begin{figure}[t]
  \centering
	\includegraphics[width=1\columnwidth]{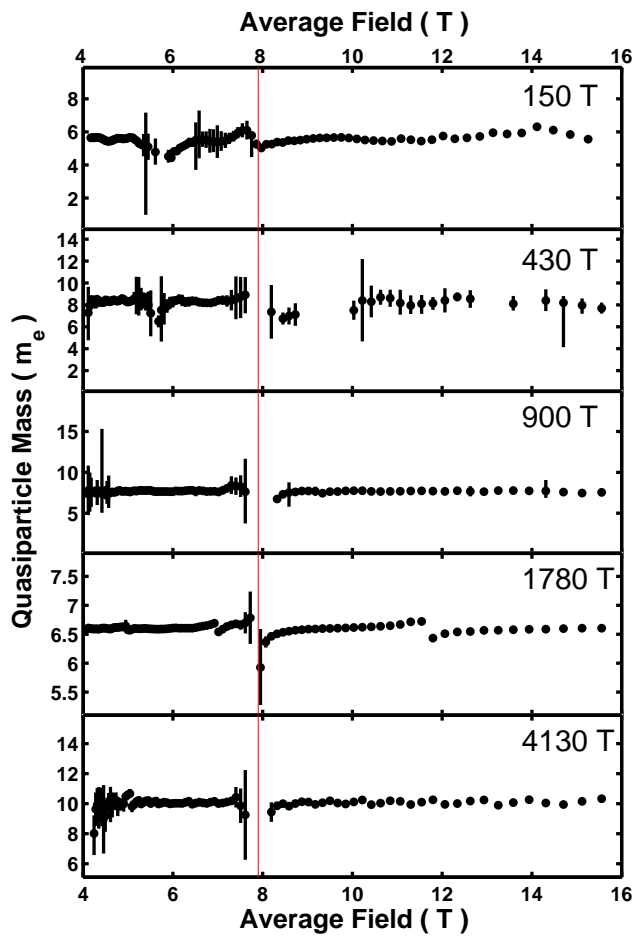}
	\caption{Quasiparticle mass of five simulated dHvA frequencies similar to those detected in our experiments, calculated using the same method as that used in Fig.~\ref{fig: FirstMassA}. The data was modelled using a constant quasiparticle mass and a scattering rate which was strongly enhanced near 7.9 T, driving the amplitude to zero at this field value, similar to the experimental observations.}
	\label{fig: MassesNonDiv2}
\end{figure}

In Fig.~\ref{fig: MassesNonDiv2} we show the mass analysis of the second data set which does not feature mass enhancements, and yields the appropriate mass values. Appendix \ref{App:B} describes how the previous incorrect result probably arose as a consequence of using a free offset as a third parameter in the Lifshitz-Kosevich fit to attempt to process data with a low signal to noise ratio. Since the current data had a much better signal to noise ratio, the use of either two or three parameter fits did not produce significantly different results.


\begin{figure}[t]
  \centering
	\includegraphics[width=1\columnwidth]{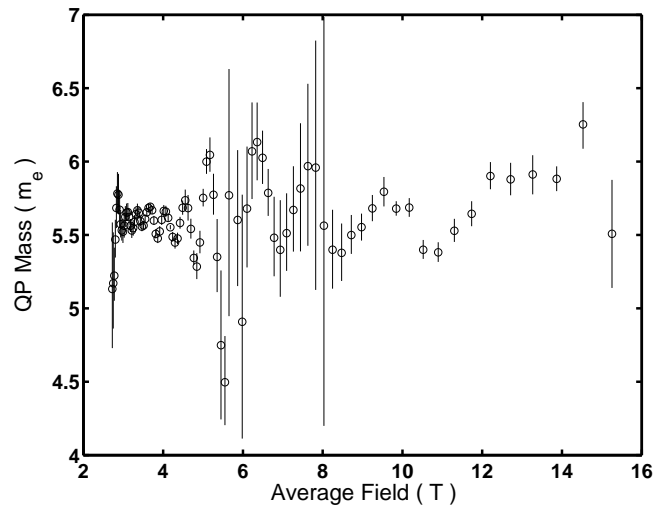}
	\caption{Quasiparticle mass extracted from the merged peak composed of the frequencies at 0.11 and 0.15~kT, originating respectively from the $\gamma_2$ and $\beta$ bands, where in the simulation, the mass of $\gamma_2$ band is strongly enhanced while that of the $\beta$ band is not. The mass enhancement is washed out by the $\beta$ band, which possesses a mass of 5.5~$m_e$.}
	\label{fig: 110TMassDiv}
\end{figure}

In a third set of tests we experimented with assigning a diverging mass individually to different frequencies, in case the entire specific heat change were due to a diverging mass on only one Fermi surface sheet. In doing so, we confirmed that if this were happening, our data analysis procedures would be sensitive to it. The only exception to this came when trying to include a diverging mass for a frequency at 0.11~kT.  Here, it is impossible to perform the field-dependent mass analysis and still resolve that component of the signal from the one at 0.15~kT.  In first harmonic dHvA, the integrated weight of the 0.11~kT component is much less than that of the 0.15~kT component, and the effect of the 0.11~kT component experiencing a rapid change of amplitude with temperature is masked by the much weaker temperature dependence of the 0.15~kT component. This is demonstrated in Fig. \ref{fig: 110TMassDiv}, which shows the result of a simulation incorporating the experimental oscillatory amplitudes in which the mass of $\gamma_2$ is enhanced by an order of magnitude between low / high fields and the transition region. The relative integrated weights of our magnetocaloric data are similar at our lowest temperature of measurement, but for that apparatus the base temperature is only 100~mK, and resolving a strongly field dependent mass for the 0.11~kT component would require lower temperature data.  We conclude, therefore, that we can neither definitely detect nor rule out a strong field dependence of the quasiparticle mass from this Fermi surface sheet on the basis of the current data.

\section{Conclusion}

The first two pieces of new information that we have presented in this paper concern the magnetic field and field angle dependence of the dHvA frequencies both below and above the metamagnetic transition fields in \TTS. Although the data contain some unusual features, it seems possible to account for these in terms of previously known effects. Field-dependent frequencies are a consequence of back-projection in the presence of itinerant metamagnetism\cite{JulianUPt3}$^,$\cite{mercure}.  The 'fragmentation' of the observed frequencies at high fields can be qualitatively accounted for by magnetic breakdown effects, although we do not have sufficient information to perform the full quantitative analysis that would be required to prove that this is the sole cause of the observations.

The finding that we observe essentially no systematic field dependence of the quasiparticle masses is more surprising. Taking a conventional viewpoint, it is tempting to assume that since the specific heat shows a large change over the same field range, some quasiparticle mass enhancement must be taking place, and hence that it must be associated with the $\gamma_2$ frequency for which we have no reliable dHvA information on the field dependence of the mass. Adopting that assumption would imply an order of magnitude change of its mass between low field and 7.5~T \footnote{At first sight it may seem that the combination of specific heat and dHvA is incompatible with previous transport data that show a factor of seven rise in the strength of $T^2$ scattering at low $T$ as the critical point is approached \cite{science1}. At zero field and without impurity scattering, the temperature dependent resistivity contribution from the $\gamma_2$ pocket, which comprises a maximum of only one fifth of the Fermi surface, would be shorted out by the other pockets. However, at finite field as well as in the limit of residual scattering dominating the resistivity a realistic calculation based on the multi-band Fermi surface of \TTS\ shows that the transport could, in principle, be accounted for by a factor of 10 increase in the $\gamma_2$ mass.}. It would be dangerous to draw a firm conclusion on this point in the absence of firm data; this aspect of our work motivates further experimental and theoretical study.

\section{Acknowledgements}

We thank A. M. Berridge, C. A. Hooley, A. G. Green and G. G. Lonzarich for informative discussions. This work was supported by the Engineering and Physical Sciences Research Council. 

\appendix

\section{Systematic errors in dHvA mass analysis \label{App:B}}

We present in this appendix a discussion of dHvA quasiparticle mass analysis methods, and how certain assumptions in the analysis method can lead to large systematic errors. In particular, we try to explain how the analysis performed in the work of Borzi $et$ $al.$ \cite{borzi} led to the apparent observation of a strong quasiparticle mass enhancement of the $\alpha_1$ and $\alpha_2$ bands (at 1.78 and 4.13~kT). 

The usual analysis method of the temperature dependence of the amplitude of dHvA oscillations consists of performing Fourier transforms of data for all temperatures over the same field range. Integrals are performed over a specific peak in the resulting power spectra, of which the square root is taken. The field dependent amplitudes are then analysed using a non-linear fit of the standard LK expression, of the form given in eq. \ref{eq:LK}. This function saturates at the lowest temperatures towards the parameter $A$ of Eq. \ref{eq:LK}, and exponentially decreases at high temperatures.

However, in quantum oscillation data which feature appreciable levels of noise compared to the signal, the integral of the power spectra never decreases to zero amplitude at high temperatures. Effectively, if we write the signal as $A \cos(F_0/B +\phi) + f(B)$, where f(B) consists of random noise, the power spectrum $W(F)$ as a function of frequency $F$ of the Fourier transformed signal can then be written as
\beq
W(F) = \big| \delta(F-F_0) + \tilde{f}(F)\big|^2,
\label{eq:Pspec}
\eeq
where $\delta(F-F_0)$ corresponds to a peak centred at $F_0$ and $\tilde{f}(F)$ is the Fourier transform of $f(B)$. At high temperatures, the peak vanishes and one is left with $W(F) = |\tilde{f}(F)|^2$, which is not zero. In the analysis of ref. [\onlinecite{borzi}] we subtracted the apparent resulting offset in the high temperature range of the temperature dependence of the dHvA amplitude by using a non-linear fit procedure involving a free offset\footnote{Private communication.}, of the form
\beq
LK(B_0,T) = A{Cm^*T/B_0 \over \sinh(Cm^*T/B_0)} + Const.,
\eeq
using three fit parameters, $A$, $m^*$ and $Const$. We will demonstrate in the following that such a procedure always leads to an overestimate of the mass $m^*$.

\begin{figure}[t]
  \centering
	\includegraphics[width=1\columnwidth]{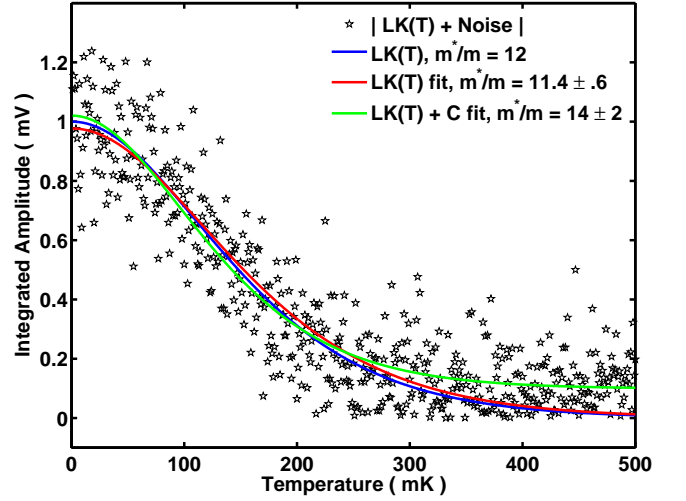}
	\caption{Example of the two types of non-linear LK fits to the data. The absolute value of the sum of the LK function (blue curve) and random gaussian noise is shown in black stars, using a quasiparticle mass of 12~$me$. To this data was fitted the LK expression using a two parameter fit (red curve), which yielded a small underestimate of $11.4\pm.6$~$m_e$. A three parameter fit was also performed, featuring a free offset (green curve), which yielded an overestimate of the mass of $14\pm2$~$m_e$.}
	\label{fig: LKNoiseFit4}
\end{figure}

\begin{figure}[t]
  \centering
	\includegraphics[width=1\columnwidth]{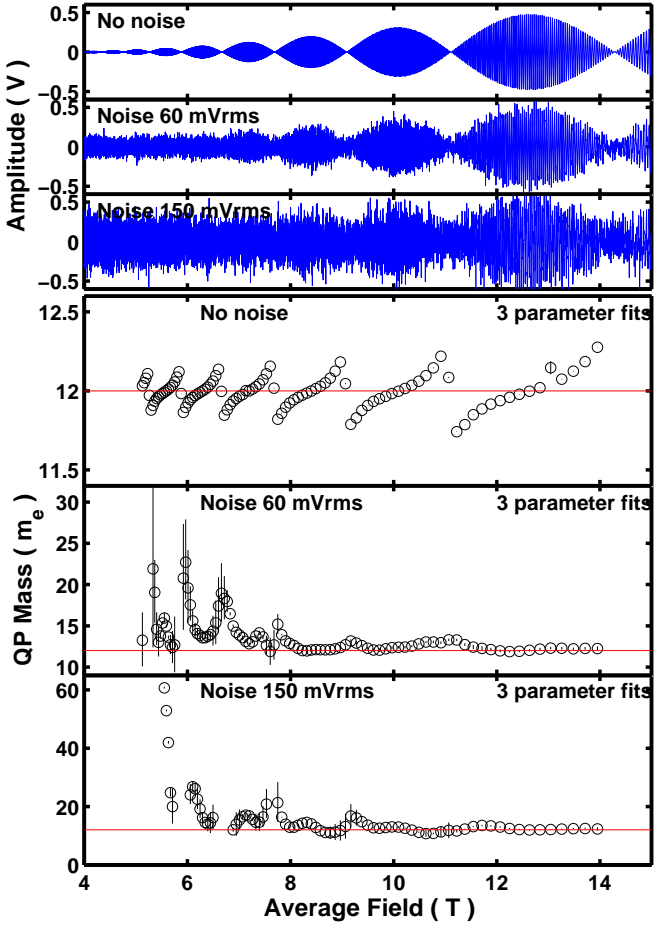}
	\caption{Simulations of dHvA data featuring a field dependent signal to noise ratio, and its mass analysis using a three parameter fit featuring a free offset. The top panel presents noiseless simulated dHvA data, featuring two frequencies of 4.20 and 4.25~kT, leading to beat patterns, a mean free path of 1~$\mu$m and a quasiparticle mass of 12~$m_e$. The following two panels show the same data with two different added noise levels. These sets of data were analysed with three parameter fits, the results of which are presented in the bottom three panels.}
	\label{fig: MassNonDivSim}
\end{figure}

Effectively, adding a free offset to the fitting procedure is equivalent to the assumption that the power spectrum is equal to $|\delta(F-F_0)|^2 + |\tilde{f}(F)|^2$, i.e. the sum of the signal and noise in frequency space. However, as stated in eq. \ref{eq:Pspec}, it is not the case, but rather, we have that
\beq
\big|\delta(F-F_0)\big|^2 + \big|\tilde{f}(F)\big|^2 \geq  \big| \delta(F-F_0) + \tilde{f}(F)\big|^2,
\eeq
and the difference between the two sides of this inequality depends on the temperature. As the noise amplitude increases, the difference between these expressions increases, and the signal becomes `buried' into the noise. Consequently, the width of the peak emerging from the noise at low temperature appears to decrease, yielding a larger mass $m^*$ when using a free offset. It is however not what happens when one uses the usual two parameter fit.

In Fig.~\ref{fig: LKNoiseFit4} we illustrate this situation. We took the normal LK distribution (blue curve), using a mass of 12~$m_e$, to which we added random gaussian noise, and took the absolute value (the square root of the power spectrum), shown in black stars. We used 500 data points in order to show that the problem does not arise due to under-sampling. We then performed the two different types of fits, with (green curve) and without (red curve) a free offset. The resulting masses are different, where the two parameter fit obtains a slight underestimate (5\%) while the three parameter fit overestimates the value by 17\%. While neither of the two methods yields the correct value, the three parameter fit always deviates more strongly, and always overestimates the value.

This effect depends strongly on the signal to noise ratio. Consequently, if the signal to noise ratio decreases gradually as a function of magnetic field, then the overestimates generated by three parameter fits usually increases gradually, giving the impression of a field varying mass. In order to demonstrate the serious nature of this problem,  we performed a simulation of typical dHvA data (similar to that of the $\alpha_2$ band), which featured beat patterns, a Dingle amplitude reduction factor and a constant level of uncorrelated gaussian distributed noise of various amplitudes:
\beq
LK(B,T)\dot D(B) \dot \bigg[\cos(2\pi F_1/B) + \cos(2\pi F_2/B) \bigg] + f(B),
\eeq
where $LK$ is the LK function, which uses a mass of 12~$m_e$, $D$ is the Dingle factor, which uses a mean free path of 1~$\mu$m and a frequency of 4.2~kT and $f(B)$ is the noise. 

In Fig.~\ref{fig: MassNonDivSim}, top three panels, we show simulated raw data, the first without noise and the following two with different noise amplitudes. The bottom three panels present the respective extracted field dependent masses when using three parameter fits. The procedure for this calculation was the same as that which was used for the experimental data, described earlier. We found strong systematic errors in both cases where noise was added to the data, which are largest at each node of the beat pattern in the region where the overall signal to noise ratio is the lowest. We performed the same analysis while using two parameter fits, not shown here, and this problem was not encountered. 

In \TTS\ dHvA data, the signal to noise ratio decreases near the metamagnetic transition, mainly due to a combination of the Dingle reduction factor and scattering which increases near the QCEP and reduces the amplitude of quantum oscillations. Consequently, it is natural that this problem could have arisen specifically in this field range. However, it did not arise in the analysis of the data reported in this paper which featured a much higher signal to noise ratio, where the resulting offset was generally close to zero. Such fits generated the same results as two parameter fits.

\bibliographystyle{apsrevNOETAL}

\end{document}